\documentclass[%
 reprint,
 amsmath,amssymb,
 aps,
prl,
]{revtex4-1}

\usepackage{graphicx}
\usepackage{dcolumn}
\usepackage{bm}
\usepackage{epstopdf}

\begin{document}

\preprint{APS/123-QED}

\title{Scalable Loading of a Two-Dimensional Trapped-Ion Array}

\author{C. D. Bruzewicz}
\email{colin.bruzewicz@ll.mit.edu}
\author{R. McConnell}
\email{robert.mcconnell@ll.mit.edu}
\author{J. Chiaverini}
 \email{john.chiaverini@ll.mit.edu}
\author{J. M. Sage}
 \email{jsage@ll.mit.edu}
\affiliation{Lincoln Laboratory, Massachusetts Institute of Technology, Lexington, Massachusetts 02420, USA}

\date{\today}

\begin{abstract}
We describe rapid, random-access loading of a two-dimensional (2D) surface-electrode ion-trap array based on two crossed photo-ionization laser beams. With the use of a continuous flux of pre-cooled neutral atoms from a remotely-located source, we achieve loading of a single ion per site while maintaining long trap lifetimes and without disturbing the coherence of an ion quantum bit in an adjacent site. This demonstration satisfies all major criteria necessary for loading and reloading extensive 2D arrays, as will be required for large-scale quantum information processing. Moreover, the already high loading rate can be increased by loading ions in parallel with only a concomitant increase in photo-ionization laser power and no need for additional atomic flux.
\end{abstract}

\pacs{Valid PACS appear here}

\maketitle
Trapped ions have the potential to form the basis of a large-scale quantum processor due to ion internal states' natural isolation from environmental disturbances and to the straightforward, high-fidelity methods developed to manipulate those states~\cite{harty2014high}. However, arrays of many ions will require site reloading when an ion is lost due to collisions or reactions with background gas species. Even in cryogenic vacuum systems with single-ion lifetimes greater than tens of hours~\cite{antohi2009cryogenic}, an array of $10^{7}$~ ions, which is a reasonable estimate of the physical qubit count in a fault-tolerant architecture~\cite{PhysRevA.79.062314,PhysRevA.86.032324}, will require continuous reloading of empty sites at an average rate of approximately 100~s$^{-1}$. In addition, ions will be lost at random locations throughout the array, necessitating random-access loading at high rates. The ion reloading process must also not lead to unacceptable levels of decoherence in nearby trapped-ion qubits. Otherwise, fault-tolerance may be compromised.

Refilling an array from loading zones at the array's edge is limited by the time required to move an ion to interior sites and requires additional complexity in the trap electrode structure to transport ions throughout the array. If ions are instead introduced into the array only at the edge to eliminate these requirements~\cite{hensinger2015architecture}, a large overhead of quantum-logical-swap operations that scales poorly with array size is accrued. Loading ions through holes in the chip \cite{:/content/aip/journal/apl/95/17/10.1063/1.3254188} may potentially be implemented with many holes near the array sites to allow rapid random access, but this would likely preclude the on-chip integration of electronic and photonic components necessary for scalable control and readout across an array. 

Here we demonstrate two-dimensional (2D) ion-trap array loading that, uniquely among implemented or proposed methods, satisfies all requirements for scalability to large numbers of ions. Using a continuous, pre-cooled, neutral atomic beam, we rapidly load sites with random access and without moving any ions in or through the array. The spatial separation of the pre-cooled atom source from the ion-trap array allows for the continuous cold-atom flux while still providing long ion lifetimes in scalable surface-electrode traps. We also show that site-specific ion loading can be accomplished while introducing, at most, a negligible amount of qubit memory error in neighboring sites. Therefore, quantum processing in other parts of the array may continue during ion replacement without additional error, allowing for fault-tolerant operation. The method demonstrated here is an enabling capability for practical operation of larger-scale trapped-ion quantum information systems. 

The cryogenic ion-trapping apparatus used in this work is shown in Fig.~1(a) and is a variation of a system used previously \cite{sage2012loading,chiaverini2014insensitivity,bruzewicz2015measurement}. In order to achieve fast ion-loading rates and long trap lifetimes, we have implemented a two-dimensional magneto-optical trap (2D-MOT) of neutral strontium operating on the $5S_{0}\!\to\!5P_{1}$ transition at 461~nm. Separating the MOT from the cryogenic ion-trapping  system by a narrow differential pumping tube  permits continuous operation of the atomic oven at 400$^{\circ}$~C with a vacuum pressure of $2\!\times\!10^{-8}$~Torr in the MOT chamber without limiting ion lifetimes by background gas collisions. The design uses stacks of permanent magnets to generate a 2D quadrupole field that vanishes along the atomic beam axis \cite{tiecke2009high,lamporesi2013compact}. An additional weak push laser beam tuned near the MOT transition accelerates the atoms towards the cryogenic ion-trapping chamber. The pre-cooled atoms travel through the differential pumping tube and holes in the windows of the 50~K and 4~K radiation shields of the ion-trapping chamber. 

\begin{figure}
\begin{center}
\includegraphics[width=\columnwidth]{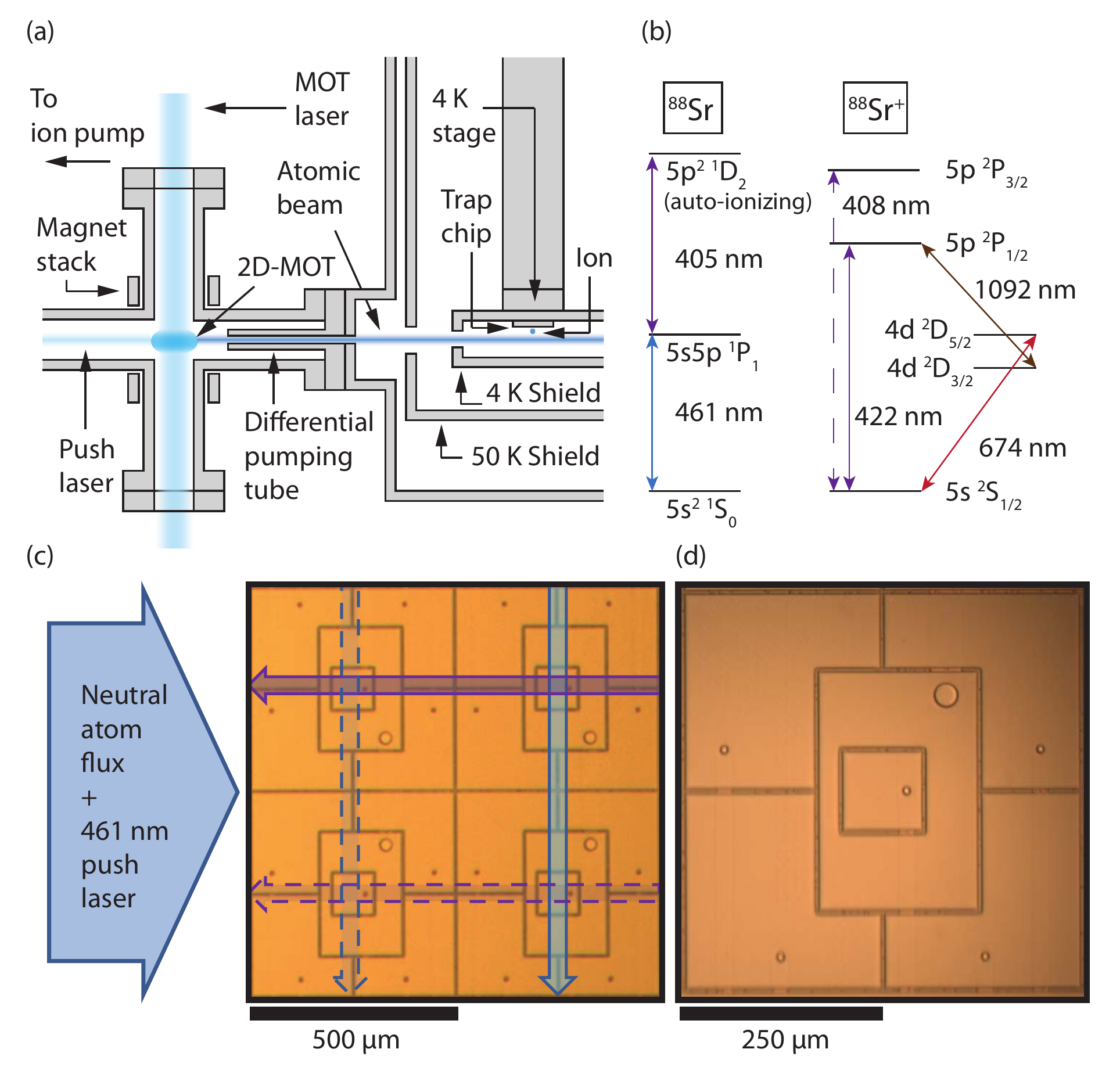}
\caption{(Color online) (a) Simplified schematic of two-dimensional magneto-optical trap apparatus. (b) Relevant transitions in neutral and singly-ionized $^{88}$Sr. Energy splittings not drawn to scale. (c) Schematic representation of the site-selective loading scheme overlaid on a micrograph of the $2\!\times\!2$ trap array. 461~nm (vertical/blue) and 405~nm (horizontal/violet) photo-ionization beams propagate orthogonally to each other such that only the site to be loaded (upper right in this case) is illuminated with both wavelengths. Dashed arrows denote paths of PI beams used to load other array sites. (d) Micrograph of an individual trap array site. The rectangular RF ring electrode sufficiently tilts the trap axes to permit efficient Doppler-cooling with a single laser beam. Electrical connections to the trap electrodes are made using the circular interlayer vias.}
\label{fig:schematic}
\end{center}
\end{figure}

Near the ion trap chip, two focused lasers propagating parallel to the trap surface produce  $^{88}\mathrm{Sr}^{+}$ ions from the cold atomic beam by means of a two-step photo-ionization (PI) process, shown schematically in Fig.~\ref{fig:schematic}. A resultant  ion is subsequently Doppler cooled and trapped 50~$\mu$m from the surface-electrode point-Paul-trap array.  

Three-dimensional confinement is provided by a time-varying radio frequency (RF) voltage applied to the rectangular ring electrodes (see Fig.~1(c)), yielding radial trapping frequencies of a few megahertz \cite{kim2010surface}.  The RF amplitude is adjusted such that only a single ion can be trapped in each site. Operating in this stability regime reduces the average trap lifetime to a few hours, but lifetimes greater than 18 hours in the presence of Doppler-cooling light have been observed at lower RF amplitude. Segmented DC electrodes inside and surrounding the RF electrode are used to adjust the location of the ion and compensate for stray electric fields. 

The trap array consists of four separate traps arranged in a $2\!\times\!2$ square geometry with an array pitch of 500~$\mu$m. Although ion-ion interactions are too small at this distance for practical multi-qubit logic gates, we expect future designs will be of a similar size with additional electrodes to permit shuttling of ions to adjacent sites when performing two-qubit gates \cite{kielpinski2002architecture}. The additional space afforded by this array pitch may permit the use of integrated photonic devices to route the large number of laser beams needed for scalable operation.

\begin{figure}
\begin{center}
\includegraphics[width=\columnwidth]{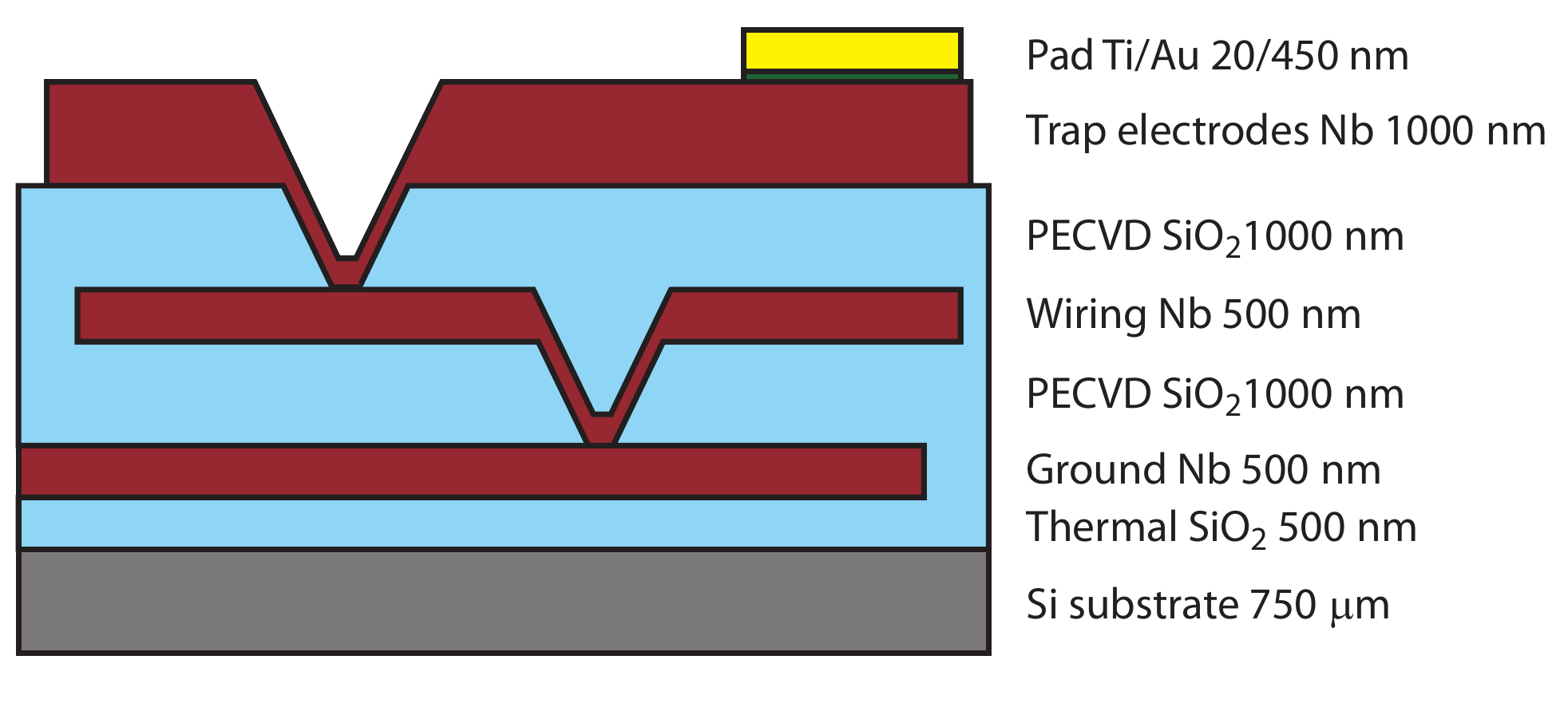}
\caption{(Color online) Multilayer stack of trap chip (not drawn to scale). Surface electrodes, wiring layer, and ground plane are made of sputtered niobium (Nb). Insulating layers are formed by plasma-enhanced chemical vapor deposited (PECVD) silicon dioxide (SiO$_{2}$).}
\label{fig:traps}
\end{center}
\end{figure}

The trap chip was fabricated using a superconducting multilayer process shown in Fig.~\ref{fig:traps}. The trap electrodes as well as the wiring layer are made from sputtered niobium and are insulated by interspersed layers of plasma-enhanced chemical vapor deposited (PECVD) silicon dioxide. An additional niobium ground plane was deposited between the wiring layer and the substrate to prevent  optically-generated charge carriers within the silicon from affecting the trap impedance \cite{mehta2014ion}. Electrical connections to the trap electrodes are made using interlayer vias contacting the wiring layer. This wiring layer is routed to gold pads at the side of the chip that are wire-bonded to a ceramic pin grid array (CPGA) chip carrier. The CPGA chip carrier is mounted on an electronic filter board that is attached to the cold head of the cryogenic system.  In this configuration, the trap reaches a steady-state temperature of $\sim\!8$~K at the highest RF amplitude used here.

To avoid additional computational overhead, scalable loading of a 2D ion-trap array must be site-selective. We achieve this by aligning the 461~nm and 405~nm lasers that drive the two-step photo-ionization orthogonally to each other such that only the chosen site to be loaded is addressed by both necessary wavelengths. We are able to quickly switch the locations of the photo-ionization lasers to address any desired site by changing the driving frequencies and deflection angles of steering acousto-optic modulators (AOM) in the laser beam paths \cite{naegerl1999laser}. In our current geometry, AOM shifts of 50-60~MHz are sufficient to achieve the necessary 500~$\mu$m beam translations between array sites. The 461~nm laser is sent through an additional AOM in a double-pass configuration that keeps the laser frequency on resonance as the steering AOM frequency is changed. The auto-ionizing transition at 405~nm is sufficiently broad that such frequency compensation is not necessary for this beam.

We determined the ion loading rates by measuring the loading probability of a chosen array site as a function of photo-ionization time. To achieve the highest loading rates, the 2D-MOT and push laser beams ran continuously such that neutral atoms were always available to be photo-ionized at the trap locations. In each trial, the PI laser beams were pulsed on for a variable time, followed by a 2~ms pulse of 422~nm Doppler-cooling light. We then waited 8~ms without Doppler-cooling to ensure that any transiently-trapped ions had left the trap.  Following this delay, we measured resonant fluorescence during a second 2~ms Doppler-cooling pulse to detect the presence of a single, stably-trapped ion. After detection, a modest positive voltage (1~V) was applied for 1~ms to the center square DC electrode to eject any trapped ion. An ion-repumping laser at 1092~nm (see Fig.~1(b)) also illuminated the array throughout the loading experiments. 

\begin{figure}
\begin{center}
\includegraphics[width=\columnwidth]{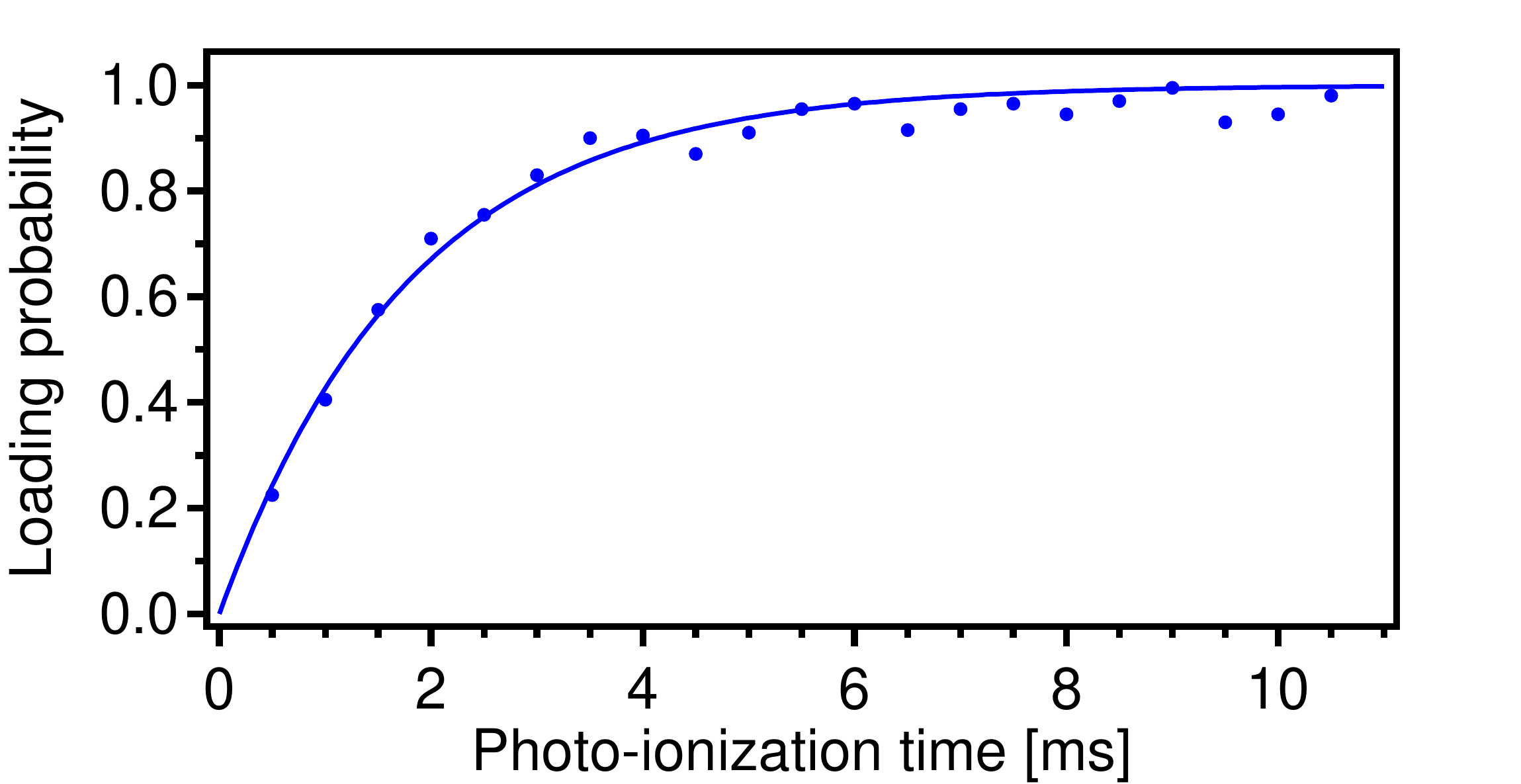}
\caption{(Color online) Loading rate of a single ion trap array site (upper right in Fig.~\ref{fig:schematic}). Each trial was repeated 200~times per point. Fit to model $P_{\mathrm{load}}=1-e^{-t/\tau_{\mathrm{load}}}$ yields a time constant $\tau_{\mathrm{load}}=1.80(6)$~ms.}
\label{fig:load}
\end{center}
\end{figure}

These trials were repeated 200 times for each PI load time to determine the loading probability. The trap loading rate was characterized by fitting the loading probability as a function of photo-ionization time to an exponential model $P_{\mathrm{load}}=1-e^{-t/\tau_{\mathrm{load}}}$. From this fit, we define an average loading rate $r=1/\tau_{\mathrm{load}}$ for the Poissonian loading process. The results of these measurements are given in Fig.~\ref{fig:load} and Table I, showing average loading rates greater than 400~s$^{-1}$  in all array sites.

\begin{table}
\renewcommand{\arraystretch}{1.3}
\begin{tabular}{l c c }
\hline
\hline
Array site & $\tau_{\mathrm{load}}$ [ms] & Loading rate [s$^{-1}$]  \\
\hline
Upper right & 1.80(6) & 560(20) \\
Upper left & 1.95(5) & 510(10) \\
Lower right & 2.37(5) & 420(10) \\
Lower left & 2.47(6) & 410(10) \\
\hline
\hline
\end{tabular}
\caption{Average loading times and rates for each array site. Values in parentheses reflect uncertainties from the model fit.}
\end{table}

In addition to loading site-selectively, it is also desirable to deterministically load a single ion into a given site. Therefore, inadvertent loading into occupied array sites must be minimized. This process can be due to the finite  size of the photo-ionization beams (1/$e^{2}$ radius $\sim\!60~\mu$m) extending into adjacent array sites or the presence of the weak 461~nm 2D-MOT push laser beam in the path of the 405~nm PI beam. We measured the incorrect-site loading probability of a particular ion trap site by looking for evidence of loading while attempting to load each adjacent site. The loading attempts consisted of 2~ms of PI and were repeated more than 50000 times for each site. This probability was found to be approximately $2\!\times\!10^{-4}$ when loading the adjacent site along the 461~nm laser beam axis (upper left in Fig.~1(c))  and $7\!\times\!10^{-4}$ when loading the adjacent site along the 405~nm laser axis (lower right in Fig.~1(c)). Under the assumption that inadvertent loading leads directly to qubit error, these probabilities are already sufficiently low for use with surface code error correction protocols \cite{fowler2009high}. 

To confirm our ability to maintain coherence of ions in all sites under the conditions necessary for rapid loading, we conducted a series of Ramsey experiments on a single trapped ion in the presence and absence of neutral atom flux as well as each of the PI laser beams. Tests of the PI beams are necessary because although only a single site is simultaneously illuminated by both beams during loading, entire rows and columns of the trap array are subject to the beams individually.  We also conducted similar Ramsey experiments while loading adjacent trap sites.  Each trial began with 1~ms of Doppler cooling, followed by resolved sideband cooling to the motional ground state of a 2.4~MHz radial trap mode.  We then drove a 6.5-$\mu\mathrm{s}$-long $\pi$/2 pulse on the narrow $5S_{1/2}\!\to\!4D_{5/2}$ transition at 674~nm. After a variable delay, we drove a second, phase-coherent $\pi$/2~pulse and measured the state of the ion. The Ramsey fringe contrast was determined by scanning the relative phase of the two $\pi$/2~pulses.

\begin{figure}
\begin{center}
\includegraphics[width=\columnwidth]{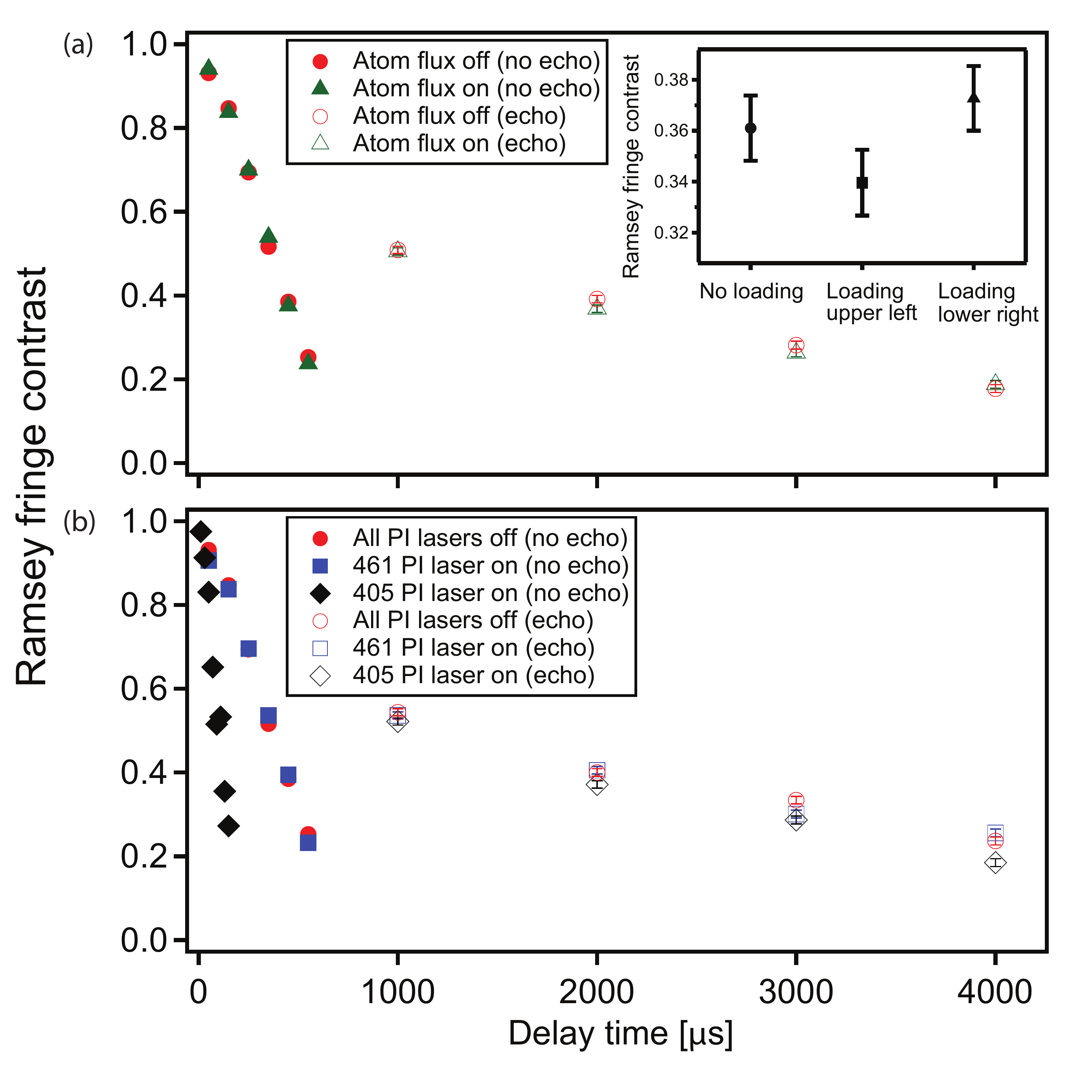}
\caption{(Color online) Coherence decay measurements performed under the conditions necessary to load site-selectively. (a) Measurements of the coherence in the presence and absence of neutral atom flux.  Inset shows the contrast while attempting to load adjacent array sites during the 2~ms Ramsey delay.  (b) Coherence in the presence and absence of the 461~nm and 405~nm PI laser beams. For the longer delay times (open markers), spin-echo $\pi$-pulses were applied at $T_{0}$, $3T_{0}$, $5T_{0}$, and $7T_{0}$, where $T_{0}\!=\!500~\mu$s. Error bars for all data points, which are comparable to the point size except in the inset, reflect uncertainty from the Ramsey fringe contrast fit with quantum projection noise propagated throughout the fitting procedure. Each phase point trial was repeated 1000 times for the primary figures and 500 times for the inset.}
\label{fig:coherence}
\end{center}
\end{figure}

The coherence time in the absence of atomic flux and PI lasers was measured to be 480~$\mu$s.  We verified that the coherence time was not limited by magnetic field fluctuations by measuring the coherence time using two transitions with different magnetic field sensitivities. The coherence times were found to be the same for both transitions, suggesting that such fluctuations did not limit our measurements. If the measured coherence decay is attributed solely to frequency fluctuations in the 674~nm laser driving the Ramsey pulses, we extract a laser linewidth of 1~kHz~\cite{sengstock}, which is consistent with direct measurements of the $S\!\to\!D$ transition. With the inclusion of appropriately timed $\pi$-pulse spin echoes to counteract slow fluctuations, we are able to extend the 1/$e$ coherence time beyond 3~ms. This technique is compatible with quantum information processing algorithms~\cite{chiaverini2004realization} and allows us to measure decoherence on time scales comparable to the ion-loading time. 

Figure 4(a) shows that the continuous flux of neutral Sr atoms had no measurable effect on the trapped  ion coherence at our current sensitivity for Ramsey delay times up to 4~ms. Collisions between atoms and the trapped ion are predicted to occur at a rate given by the product of the atomic flux $N_{\mathrm{Sr}}$ and the ion-atom collision cross-section~$\sigma$. We have estimated the atomic flux by observing fluorescence from the atomic beam at the ion trap chip location in the cryogenic trapping chamber as $N_{\mathrm{Sr}}\!\approx\!10^{8}$ cm$^{-2}\mathrm{s}^{-1}$. The relevant Langevin cross-section can be calculated using the polarizability of Sr and the approximate atom velocity $v\!\approx\!70$~m/s to yield $\sigma\!\approx\!3\!\times\!10^{-13}$~cm$^{2}$~\cite{schwartz1974measurement,chen2014neutral}. Hence, we predict a collision rate and worst-case qubit error rate per ion of approximately $3\!\times\!10^{-5}$~s$^{-1}$. Given the measured loading rates and trap lifetimes, such a collision probability would allow us to maintain arrays of $\sim\!10^{7}$~ions in the presence of continuous atomic flux, conservatively assuming that each ion-atom collision results in ion loss.

As seen in Fig.~\ref{fig:coherence}(b), the relatively weak intensity 461~nm PI beam had no effect on the trapped-ion coherence, as expected given its large detuning from all $^{88}\mathrm{Sr}^{+}$ transitions. When the ion was exposed to the 405~nm PI laser, however, a large reduction in Ramsey fringe contrast was observed. The AC Stark effect due to this much more intense beam ($I_{\mathrm{peak}}\!=\!230$~W/cm$^{2}$),  significantly shifts the $5S_{1/2}$ and $5P_{3/2}$ levels of the ion, whose transition is located near 408~nm.  This level shift caused a 60~kHz detuning of the narrow $5S_{1/2}\!\to\!4D_{5/2}$ transition that required adjusting the 674~nm laser frequency in order to drive high-fidelity Ramsey pulses. However, even when the 405~nm laser was only on during the Ramsey delays and off during the pulses, the gaussian fit coherence time without spin-echo pulses was reduced to 130~$\mu$s from 480~$\mu$s. 

We attribute the measured dephasing to low-frequency intensity noise, likely due to fluctuations in the 405~nm beam pointing. For an assumed gaussian distribution of pointing errors, the measured coherence time in the presence of the 405~nm laser corresponds to angular beam deviations at the final focusing lens of $\sim\!90$~microradians, which are consistent with measurements made using a quadrant photodiode. With the inclusion of spin-echo pulses to mitigate low frequency fluctuations, the contrast in the presence of the 405~nm laser improved dramatically. Only a slight degradation of coherence relative to the measurements without the 405~nm laser was observed for delays up to 4~ms. 

To measure trapped-ion coherence while attempting to load an adjacent trap site, the PI lasers were on for the duration of a 2~ms Ramsey experiment, which included two spin-echo $\pi$-pulses. Under these conditions, the loading probability is greater than 50\%, and the measurements were repeated 500~times per phase point.  In these experiments, we observed different behavior when loading the two sites adjacent to the upper-right site in Fig.~1(c). As seen in the inset of Fig.~\ref{fig:coherence}(a), the coherence was unaffected when loading the adjacent site along 461~nm PI laser axis (lower right in Fig.~1(c)) for 2~ms. Given that neither the 461~nm laser nor atomic flux separately reduced the contrast, this result is consistent with the expectation that the 500~$\mu$m separation between array sites is large enough that the ion-ion Coulomb interaction is too small to perturb the nearby trapped ion. When loading the site along the 405~nm PI laser axis (upper left in Fig.~1(c)), we measured coherence consistent with what was observed when the 405~nm laser was applied in the absence of atomic flux.

In conclusion, we have demonstrated a new surface-electrode ion trap loading scheme designed to site-selectively address individual traps within an array. By photo-ionizing neutral atoms from a continuous, pre-cooled atomic beam, we measured average loading rates greater than 400~s$^{-1}$ for each site of a $2\!\times\!2$ array while maintaining trap lifetimes of many hours.  Importantly, no reduction in trapped-ion Ramsey fringe contrast was observed in the presence of neutral atomic flux from our 2D-MOT system. Hence, the oven can operate continuously, affording the fastest random-access loading rates. Exposure to the intense 405~nm PI laser beam used in site-selective loading was seen to dephase the ion qubit, but this effect was almost entirely eliminated with the use of straightforward spin-echo techniques. Measurements while attempting to load adjacent trap sites showed no additional trapped-ion decoherence beyond what was caused by the 405~nm laser. Hence, with the inclusion of spin-echo pulses, we have demonstrated site-selective reloading of array sites with minimal dephasing of a nearby trapped ion. In future work, we intend to scale the arrays to accommodate more ions and to further reduce any decoherence due to the 405~nm laser by stabilizing its intensity or by implementing integrated photonic grating couplers~\cite{mehta2015coupler} to address only a single array site during loading. Additionally, grating couplers will reduce the already small probability of inadvertently loading into occupied sites and will also permit parallel loading of array sites without additional atomic flux. In this way, the loading scheme described here, which already satisfies the major criteria for scalable reloading of a large trapped-ion quantum information processor, can be refined further to yield even faster loading rates with even lower error.

\begin{acknowledgments}
We thank Peter Murphy, Chris Thoummaraj, and Karen Magoon for assistance with ion trap chip packaging. This work was sponsored by the Assistant Secretary of Defense for Research and Engineering under Air Force Contract \#FA8721-05-C-0002. Opinions, interpretations, conclusions, and recommendations are those of the authors and are not necessarily endorsed by the United States Government.
\end{acknowledgments}

\bibliography{CoherentLoading}
\end{document}